\documentclass[aps,showpacs,preprint,groupedaddress,floatfix]{revtex4}%
\usepackage{graphicx}
\usepackage{dcolumn}
\usepackage{bm}
\usepackage{amssymb}
\usepackage{amsmath}
\usepackage{amsfonts}%
\providecommand{\U}[1]{\protect\rule{.1in}{.1in}}
\providecommand{\U}[1]{\protect\rule{.1in}{.1in}}

\begin{document}
\title{Finite- to zero-range relativistic mean-field interactions}
\author{T. Nik\v si\' c}
\author{D. Vretenar}
\affiliation{Physics Department, Faculty of Science, University of Zagreb, Croatia, and }
\affiliation{Physik-Department der Technischen Universit\"at M\"unchen, D-85748 Garching, Germany}
\author{G. A. Lalazissis}
\affiliation{Department of Theoretical Physics, Aristotle University of Thessaloniki,
GR-54124, Greece}
\author{P. Ring}
\affiliation{Physik-Department der Technischen Universit\"at M\"unchen, D-85748 Garching, Germany}
\date{\today}

\begin{abstract}
We study the relation between the finite-range (meson-exchange) and zero-range
(point-coupling) representations of effective nuclear interactions in the
relativistic mean-field framework. Starting from the phenomenological
interaction DD-ME2 with density-dependent meson-nucleon couplings, we
construct a family of point-coupling effective interactions for different
values of the strength parameter of the isoscalar-scalar derivative term. In
the meson-exchange picture this corresponds to different values of the
$\sigma$-meson mass. The parameters of the isoscalar-scalar and
isovector-vector channels of the point-coupling interactions are adjusted to
nuclear matter and ground-state properties of finite nuclei. By comparing
results for infinite and semi-infinite nuclear matter, ground-state masses,
charge radii, and collective excitations, we discuss constraints on the
parameters of phenomenological point-coupling relativistic effective interaction.

\end{abstract}

\pacs{21.30.Fe, 21.60.-n, 21.60.Jz}
\maketitle


\section{\label{secI}Introduction}

The theoretical framework of nuclear energy density functionals presents the
only microscopic approach to the nuclear many-body problem that can be applied
over the whole nuclide chart, from light to superheavy nuclei, and from the
valley of $\beta$ stability to the particle drip lines. The most complete and
accurate description of ground-state properties and collective excitations of
medium-heavy and heavy nuclei is currently provided by self-consistent
mean-field (SCMF) models, based on the Gogny interaction, the Skyrme energy
functional, and the relativistic meson-exchange effective Lagrangian
\cite{BHR.03,VALR.05}. Nuclear energy density functionals are not necessarily
related to any realistic nucleon-nucleon interaction but rather represent
global functionals of nucleon densities and currents. In the mean-field
approximation the dynamics of the nuclear many-body system is represented by
independent nucleons moving in self-consistent potentials, which correspond
to the actual density (current) distribution in a given nucleus. With a small
set of universal parameters adjusted to data, SCMF models have achieved a high
level of accuracy in the description of nuclear structure.

In relativistic mean-field (RMF) theory, in particular, very successful models
have been based on the finite-range meson-exchange representation, in which
the nucleus is described as a system of Dirac nucleons coupled to exchange
mesons through an effective Lagrangian. The isoscalar scalar $\sigma$ meson,
the isoscalar vector $\omega$ meson, and the isovector vector $\rho$ meson
build the minimal set of meson fields that, together with the electromagnetic
field, is necessary for a description of bulk and single-particle nuclear
properties. In addition, a quantitative treatment of nuclear matter and finite
nuclei necessitates a medium dependence of effective mean-field interactions,
which takes into account higher-order many-body effects. A medium dependence
can either be introduced by including nonlinear meson self-interaction terms
in the Lagrangian, or by assuming an explicit density dependence for the
meson-nucleon couplings. The former approach has been adopted in the
construction of several successful phenomenological RMF interactions, for
instance, the very popular NL3~\cite{NL3}, or the more recent PK1, PK1R
\cite{PK.04} and FSUGold \cite{FSUGold} parametrizations of the effective
Lagrangian. In the latter case, the density dependence of the meson-nucleon
vertex functions can be parameterized from microscopic Dirac-Brueckner
calculations of symmetric and asymmetric nuclear matter
\cite{FLW.95,JL.98,HKL.01} or it can be phenomenological
\cite{TW.99,NVFR.02,LNVR.05}, with parameters adjusted to data on finite
nuclei and empirical properties of symmetric and asymmetric nuclear matter.

At the energy scale characteristic for nuclear binding and low-lying excited
states, meson exchange ($\sigma$, $\omega$, $\rho$, $\ldots$) is just a
convenient representation of the effective nuclear interaction. The exchange
of heavy mesons is associated with short-distance dynamics that cannot be
resolved at low energies, and therefore in each channel (scalar-isoscalar,
vector-isoscalar, scalar-isovector, and vector-isovector) meson exchange can be
replaced by the corresponding local four-point (contact) interactions between
nucleons. The self-consistent relativistic mean-field framework can be
formulated in terms of point-coupling nucleon interactions. When applied in
the description of finite nuclei, relativistic mean-field point-coupling
(RMF-PC) models \cite{MNH.92,Hoch.94,FML.96,RF.97,BMM.02} produce results that
are comparable to those obtained in the meson exchange picture. Of course,
also in the case of contact interactions, medium effects can be taken into
account by the inclusion of higher-order interaction terms, for instance,
six-nucleon vertices $(\bar{\psi}\psi)^{3}$, and eight-nucleon vertices
$(\bar{\psi}\psi)^{4}$ and $[(\bar{\psi}\gamma_{\mu}\psi)(\bar{\psi}%
\gamma^{\mu}\psi)]^{2}$, or it can be encoded in the effective couplings, i.e.
in the density dependence of strength parameters of the interaction in the
isoscalar and isovector channels. Although a number of point-coupling models
have been developed over the years, it is only more recently that
phenomenological parametrizations have been adjusted and applied in the
description of finite nuclei on a level of accuracy comparable to that of
standard meson-exchange effective interactions \cite{BMM.02}.

In a series of recent articles \cite{FKV.03,FKV.04,FKV.06}, concepts of
effective field theory and density functional theory have been used to derive
a microscopic relativistic point-coupling model of nuclear many-body dynamics
constrained by in-medium QCD sum rules and chiral symmetry. The
density dependence of the effective nucleon-nucleon couplings is determined
from the long- and intermediate-range interactions generated by one- and
two-pion exchange processes. They are computed using in-medium chiral
perturbation theory, explicitly including $\Delta(1232)$ degrees of freedom
\cite{Fri.04}. Regularization dependent contributions to the energy density of
nuclear matter, calculated at three-loop level, are absorbed in contact
interactions with constants representing unresolved short-distance dynamics.

In this work we consider the general problem of relating the finite-range
(meson-exchange) and zero-range (point-coupling) representations of effective
nuclear interactions in the relativistic mean-field framework with 
density-dependent coupling constants. In infinite nuclear matter this is, of course,
a trivial task because of constant nucleon scalar and vector densities. 
The Klein-Gordon equations of the
meson-exchange model with meson masses $m_{\phi}$ and
density-dependent couplings $g_{\phi}(\rho)$, are replaced by the corresponding 
point-coupling interaction terms with strength parameters
$g_{\phi}^{2}/m_{\phi}^{2}$. In finite nuclei, however, the problem is not so
simple. Because of the radial dependence of the densities, the expansion of the
meson propagator in terms of $1/m_{\phi}^{2}$ leads to an infinite series of
gradient terms. In practice this series has to be replaced by a finite number
of terms with additional phenomenological parameters
adjusted to low-energy data. A number of studies have shown that, both for
finite-range and for point-coupling mean-field models, the empirical data set
of ground-state properties of finite nuclei can determine only a relatively
small set of parameters in the general expansion of the effective Lagrangian
in powers of the fields and their derivatives. It is therefore not a priori
clear how to select the set of point-coupling interaction terms that will
describe structure properties at the same level of accuracy as the
meson-exchange models. An approach based on concepts of effective field theory
is only of limited use here because already at the lowest orders one finds
more parameters than can be uniquely determined from data.

The theoretical framework of meson-exchange and point-coupling relativistic
mean-field models is briefly reviewed in Sec.~\ref{secII}. The relation
between finite-range and zero-range effective interactions is investigated in
Sec.~\ref{secIII}, starting from one of the modern and most accurate
meson-exchange interactions with density-dependent vertices. Sec.~\ref{secIV}
summarizes the results and ends with an outlook for future studies.


\section{\label{secII} Theoretical framework}


\subsection{\label{secIIa}Density-dependent meson-exchange models}

In the relativistic mean-field approximation the ground-state of a nucleus is
described by the product of self-consistent solutions of the single-nucleon
Dirac equation:
\begin{equation}
\left[  \gamma_{\mu}(i\partial^{\mu}-\Sigma^{\mu}-\Sigma_{R}^{\mu}%
)-(m+\Sigma_{S})\right]  \psi=0\;, \label{Dirac-eq}%
\end{equation}
which is obtained by the variation of an effective Lagrangian with respect to
the nucleon spinor $\bar{\psi}$. In the usual $\sigma$, $\omega$, and $\rho$
meson-exchange representation, the nucleon self-energies are defined the
following relations:
\begin{align}
\Sigma_{S}  &  =g_{\sigma}\sigma\;,\label{self-vec}\\
\Sigma^{\mu}  &  =g_{\omega}\omega^{\mu}+g_{\rho}\vec{\tau}\cdot\vec{\rho
}^{\mu}+e\frac{(1-\tau_{3})}{2}A^{\mu}\;,
\end{align}
and the classical meson fields are solutions of the stationary Klein-Gordon
equations:
\begin{align}
\left[  -\triangle+m_{\sigma}^{2}\right]  \sigma(\bm{r})  &  =-g_{\sigma}%
(\rho_{v})\rho_{s}(\bm{r})\;,\label{KG-sigma}\\
\left[  -\triangle+m_{\omega}^{2}\right]  \omega^{\mu}(\bm{r})  &  =g_{\omega
}(\rho_{v})j^{\mu}(\bm{r})\;,\label{KG-omega}\\
\left[  -\triangle+m_{\rho}^{2}\right]  \vec{\rho}~^{\mu}(\bm{r})  &
=g_{\rho}(\rho_{v})\vec{j}~^{\mu}(\bm{r})\;,\label{KG-rho}\\
-\triangle A^{\mu}(\bm{r})  &  =j_{p}^{\mu}(\bm{r})\;, \label{Poisson_4}%
\end{align}
for the $\sigma$ meson, the $\omega$ meson, the $\rho$ meson (vectors in
isospin space are denoted by arrows), and the Poisson equation for the vector
potential, respectively.

When the meson-nucleon couplings $g_{\sigma}$, $g_{\omega}$, and $g_{\rho}$
explicitly depend on the nucleon (vector) density 
$\rho_{v} = \sqrt{j_{\mu}j^{\mu}}$, with $j_{\mu}=\bar{\psi}\gamma_{\mu}\psi$, 
there is an additional
contribution to the nucleon self-energy -- the rearrangement term:
\begin{equation}
\label{self-rear}\Sigma_{R}^{\mu}= \frac{j^{\mu}}{\rho_{v}}\left(
\frac{\partial g_{\omega}}{\partial\rho_{v}}\bar{\psi}\gamma^{\nu}\psi
\omega_{\nu}+ \frac{\partial g_{\rho}}{\partial\rho_{v}}\bar{\psi}\gamma^{\nu
}\vec{\tau}\psi\vec{\rho}_{\nu}+ \frac{\partial g_{\sigma}}{\partial\rho_{v}%
}\bar{\psi}\psi\sigma\right)  \;.
\end{equation}
The inclusion of the rearrangement self-energies is essential for the
energy-momentum conservation and the thermodynamical consistency of the model
(the equality of the pressure obtained from the thermodynamical definition and
from the energy-momentum tensor)~\cite{FLW.95,TW.99}.

The sources of the Klein-Gordon equations (\ref{KG-sigma}), (\ref{KG-omega}),
and (\ref{KG-rho}) are the local isoscalar and isovector densities and
currents
\begin{align}
\rho_{S}({\bm r})  &  =\sum_{k}v_{k}^{2}~\bar{\psi}_{k}({\bm r})\psi
_{k}({\bm r})\;,\label{dens_4}\\
j^{\mu}({\bm r})  &  =\sum_{k}v_{k}^{2}~\bar{\psi}_{k}({\bm r})\gamma^{\mu
}\psi_{k}^{{}}({\bm r})\;,\label{dens_5}\\
j_{TV}^{\mu}({\bm r})  &  =\sum_{k}v_{k}^{2}~\bar{\psi}_{k}({\bm r}%
)\gamma^{\mu}\tau_{3}\psi_{k}^{{}}({\bm r})\;, \label{dens_6}%
\end{align}
calculated in the \textit{no-sea} approximation: the summation runs over all
occupied states in the Fermi sea, i.e., only occupied single-nucleon states
with positive energy explicitly contribute to the nucleon self-energies.
$v_{k}^{2}$ denotes the occupation factors of single-nucleon states, and
because of charge conservation only the third component of the isovector
current contributes.

There are basically two ways to determine the medium (nucleon density)
dependence of the meson-nucleon vertex functions. In the fully microscopic
approach the symmetric and asymmetric nuclear matter Dirac-Brueckner nucleon
self-energies calculated, for instance, from a bare nucleon-nucleon
meson-exchange potential, are mapped in the local density approximation on the
mean-field self-energies that determine the single-nucleon Dirac equation
(\ref{Dirac-eq}) \cite{FLW.95,JL.98,HKL.01}. Although empirical bulk properties
of infinite nuclear matter present a very good starting point, these
pseudo-data cannot really constrain the model parameters on the level that is
necessary for an accurate description of properties of spherical and deformed
nuclei. Therefore, in a more phenomenological approach an ansatz for the
density dependence is assumed, initially guided by the microscopic
Dirac-Brueckner self-energies, but a certain number of parameters is further
fine tuned to a more accurate data base of experimental masses and charge
radii of finite nuclei. This strategy has been adopted for several very
successful semiphenomenological density-dependent interactions like, for
instance, TW-99 \cite{TW.99}, DD-ME1 \cite{NVFR.02}, DD-ME2 \cite{LNVR.05}.
For these interactions the couplings of the $\sigma$ meson and $\omega$ meson
to the nucleon are assumed to be of the form:
\begin{equation}
g_{i}(\rho) = g_{i}(\rho_{sat})f_{i}(x) \quad\text{for} \quad i=\sigma
,\omega\;, \label{gcoupl}%
\end{equation}
where
\begin{equation}
f_{i}(x) = a_{i}\frac{1+b_{i}(x+d_{i})^{2}}{1+c_{i}(x+d_{i})^{2}}
\label{fcoupl}%
\end{equation}
is a function of $x=\rho/\rho_{sat}$, and $\rho_{sat}$ denotes the nucleon
density at saturation in symmetric nuclear matter. The eight real parameters
in Eq.~(\ref{fcoupl}) are not independent. Five constraints: $f_{i}(1)=1,
f_{\sigma}^{\prime\prime}(1)= f_{\omega}^{\prime\prime}(1)$, and
$f_{i}^{\prime\prime}(0)=0$, reduce the number of independent parameters to
three. Three additional parameters in the isoscalar channel are $g_{\sigma
}(\rho_{sat}),\;g_{\omega}(\rho_{sat})$, and $m_{\sigma}$ -- the mass of the
phenomenological $\sigma$ meson. For the $\rho$ meson coupling the functional
form of the density dependence is suggested by Dirac-Brueckner calculations of
asymmetric nuclear matter~\cite{JL.98}:
\begin{equation}
g_{\rho}(\rho)=g_{\rho}(\rho_{sat})\exp[-a_{\rho}(x-1)]\;,
\end{equation}
and the isovector channel is parameterized by $g_{\rho}(\rho_{sat})$ and
$a_{\rho}$. Bare values are used for the masses of the $\omega$ and $\rho$
mesons: $m_{\omega}=783$ MeV and $m_{\rho}=763$ MeV.

The eight independent parameters are adjusted to the properties of symmetric
and asymmetric nuclear matter, binding energies, charge radii, and neutron
radii of spherical nuclei. In particular, in Ref.~\cite{LNVR.05} we have
introduced the density-dependent effective interaction DD-ME2, which has been
tested in the calculation of ground-state properties of large set of spherical
and deformed nuclei. An excellent agreement with data has been obtained for
binding energies, charge isotope shifts, and quadrupole deformations. When
used in the relativistic RPA, DD-ME2 reproduces with high accuracy data on
isoscalar and isovector collective excitations \cite{LNVR.05,PVKC.07}.

\subsection{\label{secIIb}An equivalent point-coupling model}


The basic building blocks of a relativistic point-coupling (PC) Lagrangian are
the densities and currents bilinear in the Dirac spinor field $\psi$ of the
nucleon:
\begin{equation}
\bar{\psi}\mathcal{O}_{\tau}\Gamma\psi\;, \quad\mathcal{O}_{\tau}\in
\{1,\tau_{i}\}\;, \quad\Gamma\in\{1,\gamma_{\mu},\gamma_{5},\gamma_{5}%
\gamma_{m}u,\sigma_{\mu\nu}\}\;.
\end{equation}
Here $\tau_{i}$ are the isospin Pauli matrices and $\Gamma$ generically
denotes the Dirac matrices. The interaction terms of the Lagrangian are
products of these bilinears. In principle, a general effective Lagrangian can
be written as a power series in the currents $\bar{\psi}\mathcal{O}_{\tau
}\Gamma\psi$ and their derivatives. The well-known problem, however, is that
properties of symmetric and asymmetric nuclear matter, as well ground-state
data of finite nuclei, can constrain only a small subset of interaction terms,
and the choice of this set is not unique. To build an RMF
density-dependent point coupling model that will be equivalent to the standard
meson-exchange picture, we start with the following Lagrangian:
\begin{align}
\label{PC-Lagrangian}\mathcal{L}  &  = \bar{\psi} (i\gamma\cdot\partial
-m)\psi\nonumber\\
&  - \frac{1}{2}\alpha_{S}(\hat{\rho})(\bar{\psi}\psi)(\bar{\psi}\psi) -
\frac{1}{2}\alpha_{V}(\hat{\rho})(\bar{\psi}\gamma^{\mu}\psi)(\bar{\psi}%
\gamma_{\mu}\psi) - \frac{1}{2}\alpha_{TV}(\hat{\rho})(\bar{\psi}\vec{\tau
}\gamma^{\mu}\psi) (\bar{\psi}\vec{\tau}\gamma_{\mu}\psi)\nonumber\\
&  -\frac{1}{2} \delta_{S} (\partial_{\nu}\bar{\psi}\psi) (\partial^{\nu}%
\bar{\psi}\psi) -e\bar{\psi}\gamma\cdot A \frac{(1-\tau_{3})}{2}\psi\;.
\end{align}
In addition to the free-nucleon Lagrangian and the four-fermion interaction
terms in the isoscalar-scalar, the isoscalar-vector and the isovector-vector
channels, when applied to finite nuclei the model must include the coupling of
the protons to the electromagnetic field, and derivative terms. The latter
account for leading effects of finite-range interactions that are important
for a quantitative description of nuclear density distributions, e.g. nuclear
radii. One can, of course, include derivative terms in each spin-isospin
channel, and even construct higher-order terms, but in practice data on charge
radii constrain only a single term, for instance $\delta_{S} (\partial_{\nu
}\bar{\psi}\psi) (\partial^{\nu}\bar{\psi}\psi) $. Moreover, although we assume
that medium many-body effects are encoded in the density dependence of the
four-nucleon vertex functions, a single constant parametrizes the derivative
term. This coupling has been microscopically estimated, for instance, from
in-medium chiral perturbation calculation of inhomogeneous nuclear matter
\cite{Fri.04}. It has been shown that in the relevant region of nucleon
densities $0.1$ fm$^{-3} \le\rho\le0.2$ fm$^{-3}$, the coupling of the
derivative term displays only a weak density dependence and can be
approximated by a constant value $\delta_{S} \approx- 0.7~\mathrm{fm}^{4}$.

From the variation of the Lagrangian Eq.~(\ref{PC-Lagrangian}) with respect to
$\bar{\psi}$ the single-nucleon Dirac equation (\ref{Dirac-eq}) is obtained,
with the nucleon self-energies:
\begin{align}
\Sigma^{\mu}  &  = \alpha_{V}(\rho_{v}) j^{\mu}\\
\Sigma_{TV}^{\mu}  &  = \alpha_{TV}(\rho_{v}){j}_{TV}^{\mu}\\
\Sigma_{S}  &  = \alpha_{S}(\rho_{v})\rho_{s} + \delta_{S} \triangle\rho_{s}\\
\Sigma_{R}^{\mu}  &  = \frac{1}{2}\frac{j^{\mu}}{\rho_{v}} \left\{
\frac{\partial\alpha_{S}}{\partial\rho}\rho_{s}^{2} +\frac{\partial\alpha_{V}%
}{\partial\rho}j_{\mu}j^{\mu}+ \frac{\partial\alpha_{TV}}{\partial\rho}%
{j}_{\mu{TV}} {j}_{TV}^{\mu}\right\}  \;.
\end{align}
For systems with time-reversal symmetry in the ground state, i.e. even-even
nuclei, the space components of all currents vanish ($\mathbf{j}=0$), and
because of charge conservation only the third component of the isospin current
($\tau_{3} = -1$ for neutrons and $\tau_{3} = +1 $ for protons) contributes.

For the density dependence of the vertex functions we will assume the same
functional forms as in the meson-exchange representation, i.e.
\begin{equation}
\label{scvealp}\alpha_{i}(\rho) = \alpha_{i}(\rho_{sat})f_{i}(x)
\quad\text{for} \quad i=S,V\;,
\end{equation}
where $f_{i}(x)$ is given by Eq.~(\ref{fcoupl}), $x=\rho/\rho_{sat}$, and the
exponential form for the isovector channel:
\begin{equation}
\label{vevealp}\alpha_{TV}(\rho) = \alpha_{TV}(\rho_{sat}) \exp[{-a_{TV}%
(x-1)}]\;,
\end{equation}
with the parameters $\alpha_{TV}(\rho_{sat})$ and $a_{TV}$.


\section{\label{secIII}From meson-exchange to zero-range interactions}

The link between the nucleon self-energies in the meson-exchange and
point-coupling models is obtained by expanding the propagator in the
Klein-Gordon equations (\ref{KG-sigma}-\ref{KG-rho}) $\left(  -\triangle
+m_{\phi}^{2}\right)  \phi=\mp g_{\phi}\rho_{\phi}$ ( ($-$) sign for the
scalar field and ($+$) for the vector fields, respectively). If the
density dependence of the coupling constant can be neglected, the self-energy
originating from the field $\phi$ is approximately given by
\begin{equation}
\Sigma_{\phi}=\mp g_{\phi}\phi\approx\mp\frac{g_{\phi}^{2}}{m_{\phi}^{2}}%
\rho_{\phi}\mp\frac{g_{\phi}^{2}}{m_{\phi}^{4}}\triangle\rho_{\phi}%
+\ldots\label{se-me}%
\end{equation}
This is equivalent to the self-energy of the space-isospace channel $\phi$ in
the point-coupling model
\begin{equation}
\Sigma_{\phi}=\alpha_{\phi}\rho_{\phi}+\delta_{\phi}\triangle\rho_{\phi}\;,
\label{se-pc}%
\end{equation}
if we use the following mapping
\begin{equation}
\alpha_{\phi}=\mp\frac{g_{\phi}^{2}}{m_{\phi}^{2}}\quad\text{and}\quad
\delta_{\phi}=\mp\frac{g_{\phi}^{2}}{m_{\phi}^{4}}\;. \label{PC-FR-mapping}%
\end{equation}
We note that the mapping is exact for homogeneous nuclear matter, even for
density-dependent couplings, simply because the derivative terms in
Eqs.~(\ref{se-me}) and (\ref{se-pc}) vanish.

Starting from the meson-exchange density-dependent interaction DD-ME2, 
our goal is an equivalent point-coupling
parametrization for the effective Lagrangian Eq.~(\ref{PC-Lagrangian}), with
the density dependence of the parameters defined by Eqs.~(\ref{scvealp}) and
(\ref{vevealp}).

In a first step we have adjusted the parameters of the point-coupling model to
reproduce the nuclear matter equation of state obtained with the DD-ME2
interaction. In Table \ref{TabA} we collect the corresponding parameters (set
A) of the point-coupling Lagrangian Eq.~(\ref{PC-Lagrangian}). Note that the
derivative term does not contribute in the case of homogeneous nuclear matter,
and therefore the parameter $\delta_{S}$ could not be determined. The
calculated binding energy of symmetric nuclear matter, the nucleon Dirac mass,
and the symmetry energy coincide with those obtained with DD-ME2 for all
nucleon densities, simply because the mapping from meson-exchange to
point-coupling is exact on the nuclear matter level.

To determine $\delta_{S}$, we have used the parameter set A to
calculate the surface thickness and surface energy of semi-infinite nuclear
matter for several values of the parameter $\delta_{S}$, starting from the
microscopic estimate $\approx-0.7~\mathrm{fm}^{4}$. The results are summarized
in Table \ref{TabB}. Compared to the DD-ME2 values ($t=2.108$ fm and
$a_{s}=17.72$ MeV), the PC model predicts a larger surface thickness and
considerably lower values for the surface energy. By increasing the value of
$|\delta_{S}|$ the surface energy increases, but also larger values of the
surface thickness are obtained. Note that the increase of $|\delta_{S}|$
corresponds to a reduction of the mass of the fictitious $\sigma$ meson in the
meson exchange picture (cf. Eq.~(\ref{PC-FR-mapping})), and this results in 
the increase of
the range of the interaction. From the trend shown in Table \ref{TabB}, 
obviously it is not possible to simultaneously reproduce both the surface thickness
and energy of the DD-ME2 interaction. Higher-order terms in the
expansion Eq. (\ref{se-me}) simply cannot be absorbed 
in the renormalization of the strength of the second-order 
term. When applied to finite nuclei, such a PC interaction does not
reproduce charge radii on the same level of agreement with data as DD-ME2,
which has an rms error of only 0.017 fm when compared to data on absolute
charge radii and charge isotope shifts \cite{LNVR.05}. Another possibility to
increase the surface energy (note that the sign of this quantity is opposite to
that of the bulk binding energy), is to reduce the nuclear matter binding
energy at low densities, without changing its value at saturation density.
This can be done by readjusting the parameters of the PC model. Following the
same procedure that was used in Ref.~\cite{LNVR.05} to adjust the interaction
DD-ME2, the parameters are adjusted simultaneously to properties of nuclear
matter and to binding energies, charge radii, and differences between neutron
and proton radii of spherical nuclei. For nuclear matter the following
\textquotedblleft empirical" input is used: $E/A$=-16 MeV (5\%), $\rho_{sat}%
$=0.153 fm$^{-3}$ (10\%), $K_{0}$=250 MeV (5\%), and $a_{4}$=33 MeV (5\%). The
values in parentheses correspond to the error bars used in the fitting
procedure. The binding energies of finite nuclei and the charge radii (cf.
Table \ref{TabE}) are taken with an accuracy of 0.1\% and 0.2\%,
respectively. Because of larger experimental uncertainties, the error bar
used for the neutron skin is 5\%. For the open-shell nuclei pairing
correlations are treated in the BCS approximation with empirical pairing gaps
(5-point formula).

Because the starting parameters are those of set A (cf. Table \ref{TabA}), which
is equivalent to DD-ME2 in nuclear matter, and the derivative term is included
only in the isoscalar-scalar channel, our choice is to keep fixed the
parameters of the isoscalar-vector channel, and to readjust only the three
isoscalar-scalar ($b_{s}$, $c_{s}$, $d_{s}$), and two isovector ($\alpha_{TV}%
(\rho_{sat})$, $a_{TV}$) parameters. The fit has been performed for four fixed
values of the coupling of the derivative term ranging from $\delta_{s}=-0.80$
fm$^{4}$ to $\delta_{s}=-0.86$ fm$^{4}$, and the corresponding parameters are
displayed in Table~\ref{TabC} (sets B, C, D, and E, respectively).

\begin{figure}[ptb]
\includegraphics[scale=0.6]{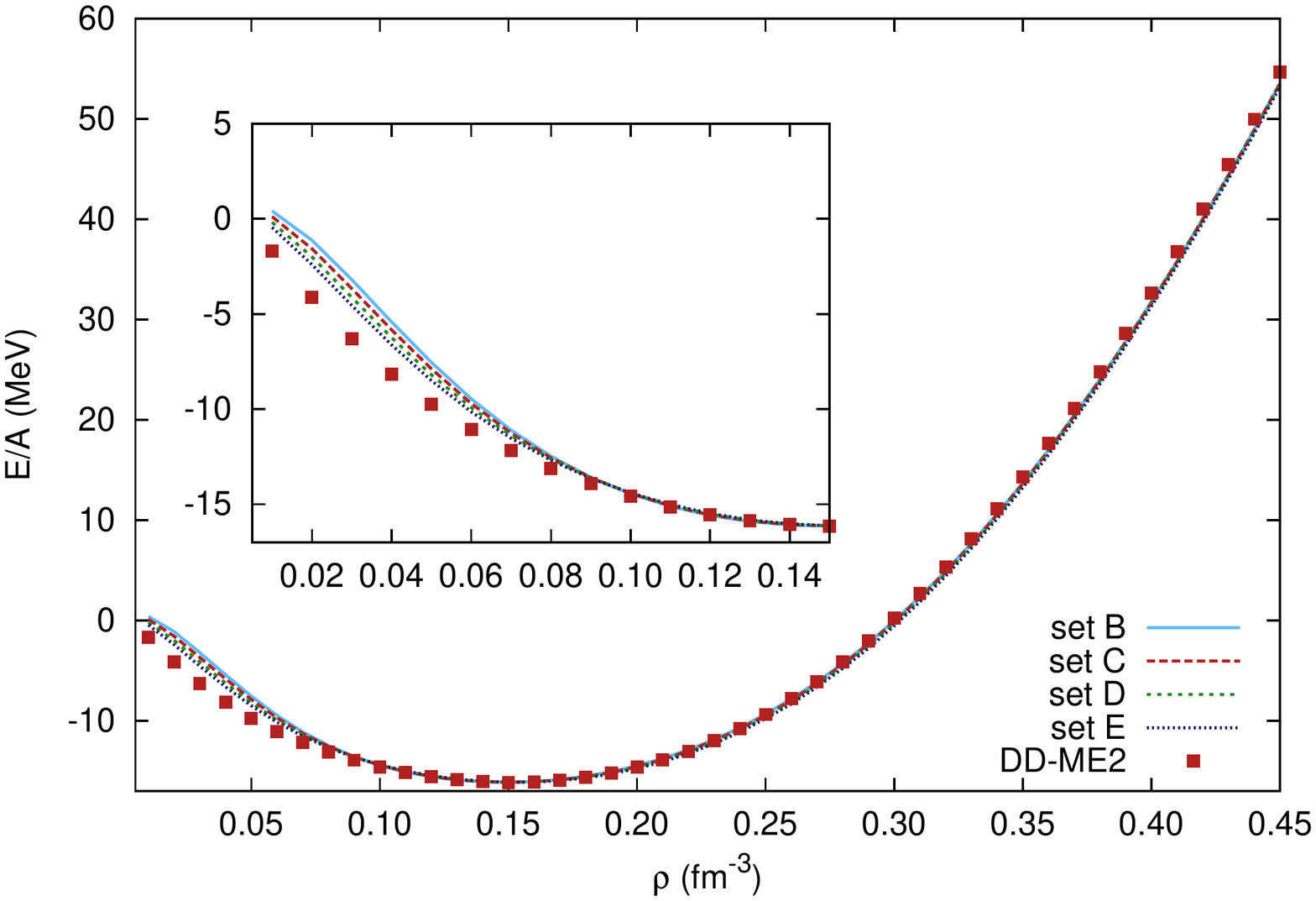} \vspace{-3cm}\caption{The equation of
state of symmetric nuclear matter (binding energy as function of nucleon
density) for the meson-exchange interaction DD-ME2, and four different
point-coupling effective interactions (cf. Table \ref{TabC}). The insert
highlights the EOS in the low-density region.}%
\label{figA}%
\end{figure}

\begin{figure}[ptb]
\includegraphics[scale=0.6]{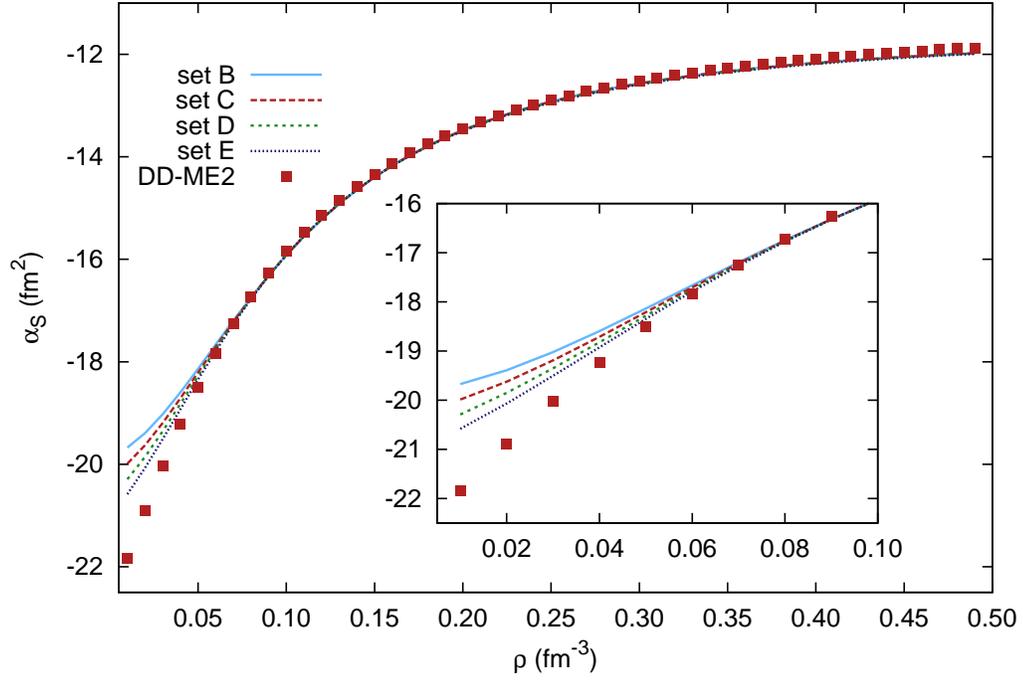} \vspace{-3cm}\caption{The density
dependence of the isoscalar-scalar coupling $\alpha_{S}$ for the
meson-exchange interaction DD-ME2, and the four point-coupling parameter sets
(cf. Table \ref{TabC}). The differences in the low-density region are
emphasized in the insert.}%
\label{figB}%
\end{figure}

\begin{figure}[ptb]
\includegraphics[scale=0.6]{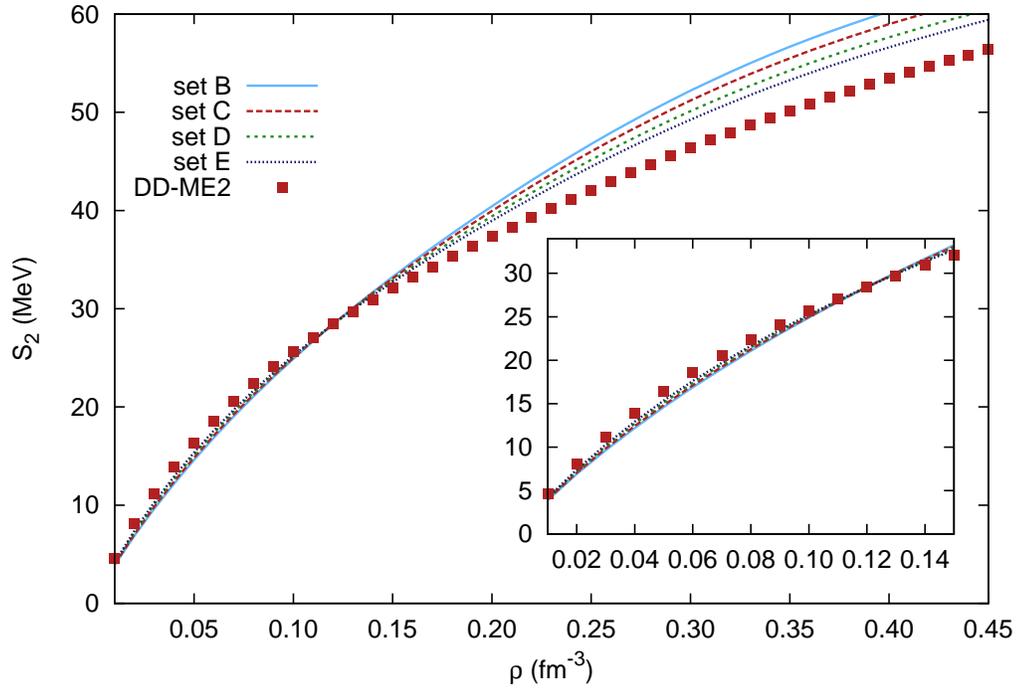} \vspace{-3cm}\caption{The symmetry
energy of nuclear matter as a function of nucleon density, for the
meson-exchange interaction DD-ME2, and the four point-coupling parameter sets
(cf. Table \ref{TabC}). The density dependence below saturation density is
shown in the insert.}%
\label{figC}%
\end{figure}

For all four parameter sets the resulting values for the surface thickness and
surface energy of semi-infinite nuclear matter (Table \ref{TabD}) are much
closer to those of DD-ME2. The corresponding binding energy curves of
symmetric nuclear matter are plotted in Fig.~\ref{figA}, in comparison with
the DD-ME2 equation of state (EOS). At nucleon densities $\rho\geq\rho_{sat}$
there is virtually no difference between the five EOS, but nuclear matter is
obviously less bound at low densities for the point-coupling effective
interactions B, C, D, and E. In the insert of Fig.~\ref{figA} we show that at
low densities the differences with respect to the DD-ME2 EOS are reduced with
the increase of the absolute strength of the derivative term. This is because
the initial value of the surface energy (calculated with the parameter set A
plus the derivative term) increases with $|\delta_{S}|$ (cf. Table \ref{TabB})
and, therefore, there is less need to compensate the larger values of
$|\delta_{S}|$ with the decrease of the nuclear matter binding energy at low
densities. The same effect can be illustrated with the density dependence of
the coupling functions of the interaction terms in the isoscalar-scalar
channel. In Fig.~\ref{figB} we display $\alpha_{S}(\rho)$ for the parameter
sets B, C, D, and E, in comparison with $\displaystyle-g_{\sigma}^{2}%
(\rho)/m_{\sigma}^{2}$ of the DD-ME2 effective interaction. For the PC
interactions the isoscalar-scalar coupling is weaker at low densities but,
with the increase of $|\delta_{S}|$, it approaches the strength function of
DD-ME2. The symmetry energy curves as functions of nucleon density for the PC
parameter sets B, C, D, and E, and the meson-exchange DD-ME2 interaction are
shown in Fig.~\ref{figC}. Here the differences are more pronounced at higher
densities above saturation, but this is simply because the ground-state data
used in the fit do not constrain the equations of state (symmetric and
asymmetric) at higher densities. At $\rho\leq\rho_{sat}$ we note that the
parabolic dependence on density is more pronounced for the DD-ME2 symmetry
energy, and that all five curves intersect at $\rho\approx0.12$ fm$^{-3}$. The
latter result is also well known in meson-exchange models for effective
interactions with different values of the symmetry energy at saturation density.

\begin{figure}[ptb]
\includegraphics[scale=0.6]{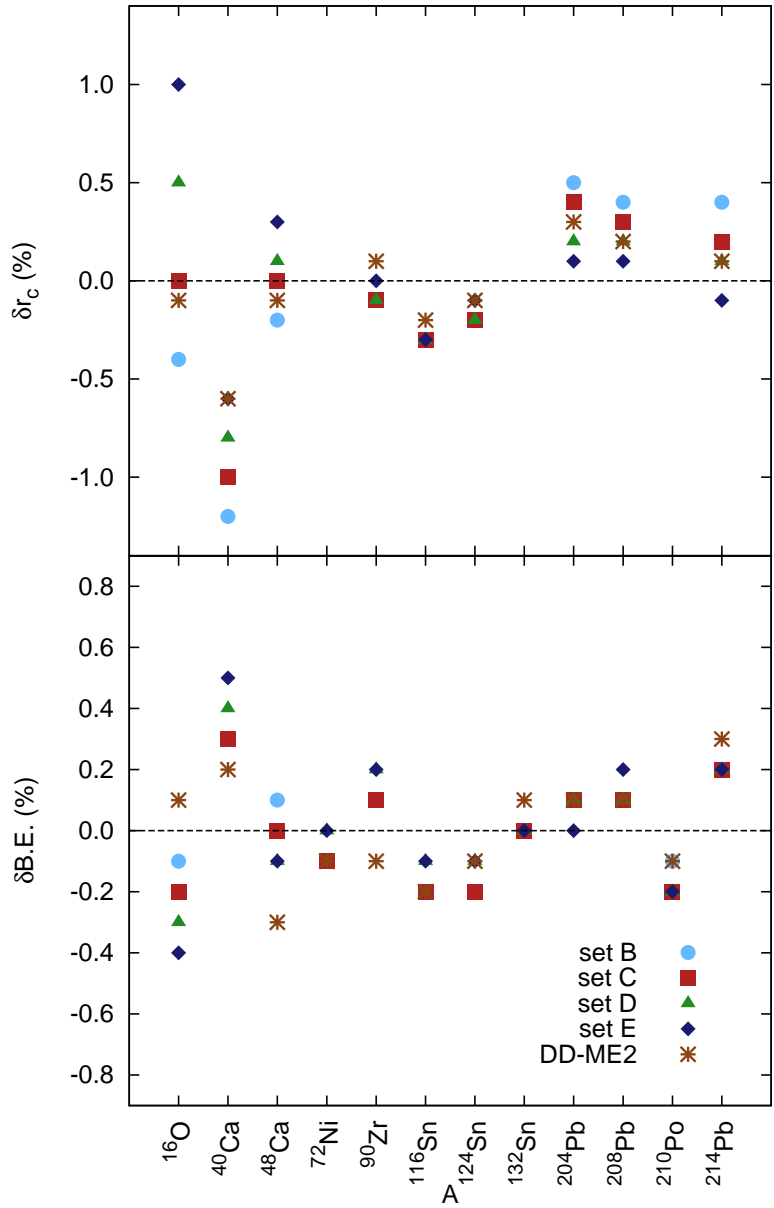}
\caption{The deviations (in percentages) between the experimental and theoretical
charge radii (upper panel), and binding energies (lower panel) of 12
spherical nuclei, calculated with the meson-exchange interaction DD-ME2, and
the four point-coupling parameter sets (cf. Table \ref{TabC}). }%
\label{figD}%
\end{figure}

With the readjustment of the parameters of the isoscalar-scalar channel
($b_{s}$, $c_{s}$, $d_{s}$) depending on the choice for $\delta_{S}$, it is
also necessary to fine-tune the two isovector parameters $\alpha_{TV}%
(\rho_{sat})$ and $a_{TV}$ to ground-state properties of the selected set of
spherical nuclei (cf. Table \ref{TabE}). In Fig.~\ref{figD} we show the
deviations in percentage between the calculated and experimental charge radii
(upper panel) and binding energies (lower panel) for the 12 spherical
nuclei that have been used to adjust the parameters of sets B, C, D, and E in
comparison with the results obtained with the meson-exchange interaction
DD-ME2. Overall, the four PC interactions reproduce data on masses and radii
on a level of accuracy comparable to that of DD-ME2: the deviations for charge
radii are generally smaller than $0.5\%$, and the agreement of binding
energies with data is better than $0.2\%$. An exception are the two lightest
nuclei $^{16}$O and $^{40}$Ca, which are notoriously difficult to describe in
a mean-field approach without including additional long-range correlations.
From the results shown in Fig.~\ref{figD} we note especially the parameter set
C for which the deviations are very close to those of DD-ME2.

\begin{figure}[ptb]
\includegraphics[scale=0.6]{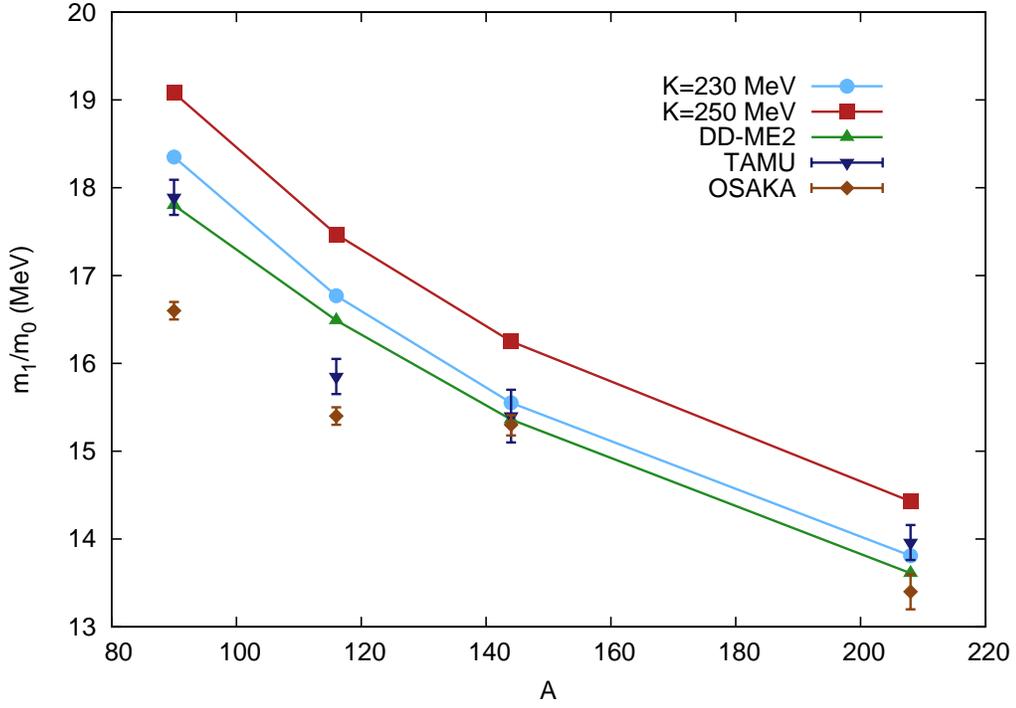} \vspace{-3cm}\caption{The RQRPA results
for the ISGMR centroid energies $(m_{1}/m_{0})$ of $^{90}$Zr, $^{116}$Sn,
$^{144}$Sm, and $^{208}$Pb, calculated with the relativistic DD-ME2
meson-exchange interaction ($K_{nm}=251$ MeV), and with two point-coupling
effective interactions: set F ($K_{nm}=250$ MeV) and set G ($K_{nm}=230$ MeV).
The theoretical results are compared with data from the
TAMU~\cite{You.97,You.99,You.04} and Osaka~\cite{Uch.03,Uch.04} compilations.
}%
\label{figE}%
\end{figure}

Better results can be obtained if the constraint on the strength of the
derivative term is released and $\delta_{S}$ is treated as a free adjustable
parameter. In the meson-exchange picture this corresponds to fitting the mass
of the $\sigma$ meson, which has been standard practice in RMF models. We have
thus refitted the parameters $b_{s}$, $c_{s}$, $d_{s}$, $\alpha_{TV}$,
$a_{TV}$ and $\delta_{S}$ on the same set of data, and the resulting best-fit
parameters are denoted as set F in Table \ref{TabC}. We note that the values
of the parameters of set F are, in fact, between those of sets C and D, i.e.
the two parametrizations which produce best results in comparison with data
(cf. Fig.~\ref{figD}). This is also true for the additional free parameter
$\delta_{S}$ for which the optimal unconstrained value is $-0.8342$, compared
to $-0.82$ (set C) and $-0.84$ (set D). Compared to the parameter sets B, C,
D, and E, the effective interaction -- set F produces results that are in
slightly better agreement with data, but an unexpected problem has appeared
when we checked the corresponding relativistic (quasiparticle) random-phase
approximation R(Q)RPA \cite{NVR.02,Paa.03} results for the excitation energies
of collective excitations -- giant resonances. Although for all five sets the
calculated excitation energies of the isoscalar quadrupole and isovector
dipole resonances in spherical nuclei are in very good agreement with data, on
the level of accuracy comparable to that of DD-ME2 \cite{LNVR.05}, the
excitation energies of the isoscalar giant monopole resonances (ISGMR) are
systematically above the experimental values. This is illustrated in
Fig.~\ref{figE} where we compare the mass dependence of the calculated ISGMR
centroid energies ($m_{1}/m_{0}$) with the data from the TAMU
\cite{You.97,You.99,You.04} and Osaka \cite{Uch.03,Uch.04} compilations for
medium-heavy and heavy spherical nuclei. Although the excitation
energies calculated with DD-ME2 (nuclear matter compression modulus
$K_{nm}=251$ MeV) are in very good agreement with data, the values obtained
with the parameter set F ($K_{nm}=250$ MeV) are consistently higher over the
whole mass range from $^{90}$Zr to $^{208}$Pb. This indicates that the surface
incompressibility of the point-coupling model is higher than that of the
corresponding meson-exchange model. It also means that to reproduce
the data on ISGMR, a relativistic point-coupling interaction should have a
nuclear matter compression modulus below 250 MeV. This difference is a very
interesting result, because in the relativistic framework with meson-exchange
interactions the interval of allowed values for K$_{nm}$ is rather narrow. In
particular, in a recent relativistic RPA analysis based on modern effective
Lagrangians with explicit density dependence of the meson-nucleon vertex
functions, we have shown that only effective interactions with K$_{nm}%
=250-270$ MeV reproduce the experimental excitation energies of ISGRM in
medium-heavy and heavy nuclei and that K$_{nm}\approx250$ MeV represents the
lower limit for the nuclear matter compression modulus of relativistic
mean-field interactions \cite{VNR.03}. This result has also been confirmed 
by the continuum relativistic RPA analysis of isoscalar monopole strength in 
$^{90}$Zr and $^{208}$Pb of Ref.~\cite{Pie.04}. By using RMF effective Lagrangians 
with nonlinear meson self-interactions,  the compression modulus of symmetric 
nuclear matter has been constrained to the range
$K_{nm} = 248 \pm 8$ MeV. 

The results shown in Fig.~\ref{figE}, however, point to a lower
value for the nuclear matter incompressibility of our point-coupling
interactions. We have thus readjusted the six parameters $b_{s}$, $c_{s}$,
$d_{s}$, $\alpha_{TV}$, $a_{TV}$ and $\delta_{S}$ on the same set of data, but
with the constraint on the nuclear matter compression modulus $K_{nm}=230$
MeV. The parameters of this interaction, denoted set G, are listed in the last
column of Table \ref{TabC}, and the resulting R(Q)RPA results for the ISGMR
centroids are shown in Fig.~\ref{figE}. The excitation energies are now much
closer to the results obtained with DD-ME2, and the comparison with data shows 
that the point-coupling interaction could even be adjusted to a slightly 
lower value of $K_{nm}$. Now a compression modulus 
$K_{nm} \approx 230$ MeV is within the range of values deduced  
from nonrelativistic self-consistent Skyrme-Hartree-Fock  plus RPA
calculations \cite{CVG.04,Col.04}. Because Skyrme forces belong to the class 
of point-coupling (contact) interactions, it would be tempting to attribute the 
well-known puzzle of different ranges of $K_{nm}$ used by standard 
relativistic mean-field interactions ($250 - 270$ MeV \cite{VNR.03,Pie.04}), 
and nonrelativistic Skyrme interactions ($220 - 235$ MeV \cite{CVG.04,Col.04}), 
to the fact that the former are of finite-range type (effective meson-exchange forces). 
However, we note that on one hand the highly successful non-relativistic finite-range 
Gogny forces also require $K_{nm} \leq  230$ MeV \cite{BBD.95a} and, 
on the other hand, it seems to be possible to construct Skyrme forces with 
$K_{nm} \approx 250$ MeV  \cite{Col.04}. Therefore, even if the present 
analysis does not resolve the puzzle of different nuclear matter compression 
moduli that have been used by relativistic and non-relativistic mean-field models, 
it shows that the gap can be further narrowed, if not completely closed, 
by exploring a new class of point-coupling relativistic interactions.

For completeness, in Table \ref{TabE} we list the 
results obtained with the parameter set G for the binding
energies, charge radii, and differences between radii of neutron and proton
density distributions of the set of 12 spherical nuclei, in comparison
with the experimental values. The agreement between the calculated values and
data is rather good and, except for $^{16}$O and $^{40}$Ca, on the same level
of accuracy as DD-ME2 (cf. Table II of Ref.~\cite{LNVR.05}). One has to keep
in mind, however, that in the present analysis all the point-coupling
interactions were based on DD-ME2 and that we have only readjusted the
isoscalar-scalar and isovector-vector channels, whereas the parameters of the
isoscalar-vector interaction were kept fixed on the values obtained by mapping
the DD-ME2 parameters in nuclear matter. An even better agreement with data
could probably be obtained by an unconstrained fit of all interaction terms of
a density-dependent point-coupling interaction.

\section{\label{secIV}Conclusions}

Models based on the self-consistent relativistic mean-field approximation have
become a standard tool for the description of ground-state properties and
collective excitations in spherical and deformed medium-heavy and heavy
nuclei. Most applications have been based on the standard meson-exchange
representation of the effective nuclear interaction, with the medium
dependence determined either from microscopic Dirac-Brueckner calculations in
nuclear matter, or in a phenomenological way by adjusting the meson-nucleon
couplings to nuclear matter and ground-state properties of finite nuclei.

Effective field theory arguments suggest that, at the low-energy scale of
nuclear structure, the picture of heavy-meson exchange associated with the
largely unknown short-distance dynamics can be effectively replaced by local
four-point (contact) interactions between nucleons. This is the basis of
relativistic point-coupling models that, in principle, must provide a
description of nuclear many-body dynamics at the level of accuracy of
phenomenological meson-exchange mean-field effective interactions. Although the
relation between the two representations is straightforward in nuclear matter,
the situation is considerably more complicated in the description of finite
nuclei. Already at the lowest order there are more point-coupling interaction
terms in the four spin-isospin channels than can be constrained by available
low-energy data. As in the meson-exchange picture one can, of course, resort
to a microscopic description of the in-medium effective nucleon-nucleon
interaction based, for instance, on chiral effective field theory. The
resulting nucleon self-energies in nuclear matter can then be mapped in the
local density approximation on the corresponding terms of a relativistic
nuclear energy density functional. However, in a completely
phenomenological approach, one makes an empirical choice of interaction terms
and adjusts the parameters to nuclear matter and data on finite nuclei. 
Both approaches require the same number of adjustable parameters 
that have to be fine tuned to low-energy data. 

In this work we have considered the mapping of a highly successful
phenomenological finite-range interaction with density-dependent meson-nucleon
couplings (DD-ME2) on the zero-range (point-coupling) relativistic mean-field
framework. We have constructed a set of point-coupling effective interactions
for different values of the strength parameter of the isoscalar-scalar
derivative term. In the meson-exchange representation this corresponds to
different values of the mass of the phenomenological $\sigma$ meson. 
The parameters of the density dependence of the isoscalar-scalar and
isovector-vector interaction terms have then been fine-tuned to nuclear matter
and ground-state properties of finite nuclei. In a comparison of the
corresponding results for infinite and semi-infinite nuclear matter,
ground-state masses, charge radii, and collective excitations, we have
discussed constraints on the point-coupling relativistic effective
interactions. It has been shown that to achieve a satisfactory
description of surface properties of semi-infinite matter and nuclear charge
radii, this class of point-coupling interactions must produce a slightly
softer nuclear matter equation of state, compared to the ones
that characterize meson-exchange interactions. Another interesting result is
that, to reproduce the excitation energies of isoscalar giant
monopole resonances, this type of point-coupling interaction requires a
nuclear matter compression modulus of $K_{nm}\approx 230$ MeV, considerably lower
than the values typically used for finite-range meson-exchange relativistic
interactions, and within the range of values used by modern nonrelativistic 
Skyrme interactions.
\bigskip\bigskip\newline\leftline{\bf ACKNOWLEDGMENTS} This work was
supported in part by MZOS - project 1191005-1010, by the DFG research cluster
\textquotedblleft Origin and Structure of the
Universe\textquotedblright, by the Alexander von Humboldt Stiftung, 
by Pythagoras II - EPEAEK II and EU project 80861.

\newpage
\newpage
\newpage
\newpage
\newpage
\newpage\begin{table}[ptb]
\caption{The parameters of the point-coupling effective interaction (set A)
that, on the nuclear matter level, is completely equivalent to the
meson-exchange interaction DD-ME2.}%
\label{TabA}
\begin{center}
\bigskip%
\begin{tabular}
[c]{lc}\hline\hline
& Set A\\\hline
$\alpha_{S}$ (fm$^{2}$) & -14.3275\\
$b_{S}$ & 0.9119\\
$c_{S}$ & 2.1127\\
$d_{S}$ & 0.3882\\
$\delta_{S}$ (fm$^{4}$) & \\
$\alpha_{V}$ (fm$^{2}$) & 10.7963\\
$b_{V}$ & 0.7648\\
$c_{V}$ & 1.8199\\
$d_{V}$ & 0.4115\\
$\alpha_{TV}$ (fm$^{2}$) & 0.9076\\
$a_{TV}$ & 1.1294\\\hline\hline
\end{tabular}
\end{center}
\end{table}
\newpage\begin{table}[ptb]
\caption{The surface thickness and surface energy of semi-infinite nuclear
matter, calculated with the point-coupling effective interaction (parameter
set A ) for various values of the strength $\delta_{s}$ of the derivative
term, in comparison with the values predicted by the DD-ME2 meson-exchange
interaction.}%
\label{TabB}
\begin{center}
\bigskip%
\begin{tabular}
[c]{lccccc}\hline\hline
$\delta_{S}$ (fm$^{4}$) & $t$ (fm) & $a_{s}$ (MeV) &  &  & \\\hline
-0.76 & 2.125 & 15.32 &  &  & \\
-0.78 & 2.157 & 15.57 &  &  & \\
-0.80 & 2.189 & 15.82 &  &  & \\
-0.82 & 2.221 & 16.06 &  &  & \\
-0.84 & 2.254 & 16.29 &  &  & \\
-0.86 & 2.286 & 16.52 &  &  & \\\hline
DD-ME2 & 2.108 & 17.72 &  &  & \\\hline\hline
\end{tabular}
\end{center}
\end{table}
\begin{table}[ptb]
\caption{The parameters of the isoscalar-scalar and isovector-vector channels
of point-coupling effective interactions, adjusted to the nuclear matter
equation of state, symmetry energy, and ground-state properties of finite
nuclei. See text for the description.}%
\label{TabC}
\begin{center}
\bigskip%
\begin{tabular}
[c]{lccccccc}\hline\hline
& set A & set B & set C & set D & set E & set F & set G\\\hline
$b_{S}$ & 0.9119 & 1.0277 & 1.0097 & 1.0010 & 1.0011 & 1.0028 & 1.1030\\
$c_{S}$ & 2.1127 & 1.7692 & 1.7839 & 1.8169 & 1.8674 & 1.8056 & 1.9386\\
$d_{S}$ & 0.3882 & 0.0674 & 0.1074 & 0.1452 & 0.1796 & 0.1344 & 0.0907\\\hline
$\delta_{S}$ (fm$^{4}$) &  & -0.80 & -0.82 & -0.84 & -0.86 & -0.8342 &
-0.8445\\\hline
$\alpha_{TV}$ (fm$^{2}$) & 0.9076 & 0.9883 & 0.9764 & 0.9618 & 0.9500 &
0.9658 & 0.9765\\
$a_{TV}$ & 1.1294 & 0.7059 & 0.7663 & 0.8368 & 0.8992 & 0.8166 &
0.8439\\\hline\hline
\end{tabular}
\end{center}
\end{table}
\newpage\begin{table}[ptb]
\caption{The surface thickness and surface energy of semi-infinite nuclear
matter, calculated with the point-coupling effective interactions: parameter
sets B, C, D, and E (cf. Table \ref{TabC}).}%
\label{TabD}
\begin{center}
\bigskip%
\begin{tabular}
[c]{lccccc}\hline\hline
& $t$ (fm) & $a_{s}$ (MeV) &  &  & \\\hline
set B & 2.015 & 17.925 &  &  & \\
set C & 2.069 & 17.856 &  &  & \\
set D & 2.126 & 17.780 &  &  & \\
set E & 2.184 & 17.717 &  &  & \\\hline\hline
\end{tabular}
\end{center}
\end{table}
\begin{table}[ptb]
\caption{The total binding energies ($B.E.$), charge radii $r_{c}$, and the
differences between the radii of neutron and proton density distributions
$r_{np}=(r_{n}-r_{p})$, used to adjust the parameters of the point-coupling
effective interactions. The values calculated with the parameter set G (cf.
Table \ref{TabC}) are compared with data (values in parentheses). In the last
three columns the corresponding deviations $dE$, $dr_{c}$, and $dr_{np}$ (all
in percentages) are included.}%
\label{TabE}
\begin{center}
\bigskip%
\begin{tabular}
[c]{lcccccc}\hline\hline
Nucleus & $B.E.$ (Mev) & $r_{c}$ (fm) & $r_{n}-r_{p}$ (fm) & $dE$ & $dr_{c}$ &
$dr_{np}$\\\hline
$^{16}$O & 127.21 (127.62) & 2.74 (2.73) & -0.03 & -0.3 & 0.5 & \\
$^{40}$Ca & 343.77 (342.05) & 3.46 (3.48) & -0.05 & 0.5 & -0.8 & \\
$^{48}$Ca & 415.76 (415.99) & 3.49 (3.48) & 0.19 & -0.1 & 0.1 & \\
$^{72}$Ni & 612.74 (613.17) & 3.92 & 0.31 & -0.1 &  & \\
$^{90}$Zr & 785.70 (783.89) & 4.27 (4.27) & 0.08 & 0.2 & -0.1 & \\
$^{116}$Sn & 987.90 (988.68) & 4.61 (4.63) & 0.12 (0.12) & -0.1 & -0.3 & 0.0\\
$^{124}$Sn & 1048.88 (1049.96) & 4.67 (4.67) & 0.21 (0.19) & -0.1 & -0.2 &
12.3\\
$^{132}$Sn & 1102.66 (1102.86) & 4.72 & 0.25 & 0.0 &  & \\
$^{204}$Pb & 1609.11 (1607.52) & 5.50 (5.49) & 0.17 & 0.1 & 0.3 & \\
$^{208}$Pb & 1639.39 (1636.45) & 5.52 (5.51) & 0.19 (0.20) & 0.2 & 0.2 &
-6.5\\
$^{214}$Pb & 1659.97 (1663.29) & 5.57 (5.56) & 0.24 & -0.2 & 0.1 & \\
$^{210}$Po & 1649.71 (1645.23) & 5.55 & 0.17 & 0.3 &  & \\\hline\hline
\end{tabular}
\end{center}
\end{table}

\bigskip
\end{document}